# Imaging and Control of Ferromagnetism in a Polar Antiferromagnet

X. Renshaw Wang[1†]*, C. J. Li[2,3†], W.M. Lü[2], T. R. Paudel[4], D. P. Leusink[1], M. Hoek[1], Nicola Poccia[1], Arturas Vailionis[5,6], T. Venkatesan[2,3,7,8], J. M. D. Coey[2,9], E. Y. Tsymbal[4], Ariando[2,7] and H. Hilgenkamp[1]

[1]MESA+ Institute for Nanotechnology, University of Twente, P.O. Box 217, 7500 AE Enschede, The Netherlands
[2]NUSNNI-Nanocore, National University of Singapore, 117411 Singapore
[3]National University of Singapore Graduate School for Integrative Sciences and Engineering (NGS), National University of Singapore, 117456 Singapore
[4]Department of Physics and Astronomy & Nebraska Center for Materials and Nanoscience, University of Nebraska, Lincoln, Nebraska 68588, USA
[5]Stanford Synchrotron Radiation Laboratory, SLAC National Accelerator Laboratory, 2575 Sand Hill Road, Menlo Park, California 94025, United States
[6]Geballe Laboratory for Advanced Materials, Stanford University, 476 Lomita Mall, Stanford, California 94305, United States
[7]Department of Physics, National University of Singapore, 117542 Singapore
[8]Department of Electrical and Computer Engineering, National University of Singapore, 117576 Singapore
[9]Department of Pure and Applied Physics, Trinity College, Dublin 2, Ireland

[†]These authors contributed equally to this work
*email: wang.xiao@utwente.nl

**Atomically sharp oxide heterostructures often exhibit unusual physical properties that are absent in the constituent bulk materials. The interplay between electrostatic boundary conditions, strain and dimensionality in ultrathin epitaxial films can result in monolayer-scale transitions in electronic or magnetic properties. Here we report an atomically sharp antiferromagnetic-to-ferromagnetic phase transition when atomically growing polar antiferromagnetic $LaMnO_3$ (001) films on $SrTiO_3$ substrates. For a thickness of five unit cells or less, the films are antiferromagnetic, but for six unit cells or more, the $LaMnO_3$ film undergoes a phase transition to a ferromagnetic state over its entire area, which is visualized by scanning superconducting quantum interference device microscopy. The transition is explained in terms of electronic reconstruction originating from the polar nature of the $LaMnO_3$ (001) films. Our results demonstrate how new emergent functionalities can be visualized and engineered in atomically thick oxide films at the atomic level.**

Modern thin film deposition techniques enable the synthesis of complex oxide thin films with unit cell (uc) level control over the thickness. Remarkably sharp phase transitions have been discovered in several systems upon increasing film thickness [1-6]. The most prominent example is the 2-dimensional electron gas formed between insulating thin films of $LaAlO_3$ and insulating $TiO_2$-terminated $SrTiO_3$ (STO) substrates, which occurs at a critical $LaAlO_3$ thickness of 4 uc [2]. The possibility to select a different electronic/magnetic phase by adding a single layer of perovskite unit cells, with a typical lattice parameter of about 4 Å, offers tantalizing opportunities for nanostructured electronic/spin devices.

Since various interesting properties have been demonstrated in LaMnO$_3$ (LMO) bulk, thin films and multilayers, ranging from the occurrence of orbital waves to its use as a catalyst for water splitting [7-11], LMO is an ideal candidate for control of its functionalities. LMO is a Mott insulator with an orthorhombic structure, based on a √2a0, √2a0, 2a0 unit cell where a0 ≈ 3.9 Å is the elementary perovskite unit cell parameter. In stoichiometric LMO, La and Mn are both 3+ ions. Therefore, LMO is a polar material which contains alternatively charged (LaO)$^{1+}$ and (MnO$_2$)$^{1-}$. Mn$^{3+}$, with electronic configuration $t_{2g}^3 e_g^1$ and spin S = 2, is a Jahn-Teller ion. If LMO had a perfect cubic perovskite structure one might expect a conducting ground state due to the mobility of the unpaired electron in the degenerate $e_g$ band. However the $e_g$ orbital degeneracy is lifted by the Jahn-Teller effect, and distorted MnO$_6$ octrahedra line up with alternating long and short Mn-O bonds in the *a-b* plane, leading to orbital ordering, which contributes to the electron localization [12]. As a result of superexchange coupling [13], the compound is an A-type antiferromagnet, with Mn spins ferromagnetically aligned in-plane, and alternate planes aligned antiferromagnetically. The Néel temperature is ~ 140 K. In bulk, the antiferromagnetism can be slightly canted to produce a weak ferromagnetism of ~ 0.18 μ$_B$ per uc [14-16], which is attributed to the antisymmetric Dzyaloshinskii-Moriya interaction introduced by rotation of the MnO$_6$ octahedra. In thin films, ferromagnetism accompanied with insulating behavior is often observed, with a Curie temperature of ~ 115 K [17]. The origin of this ferromagnetism is still unclear, but in addition to the above-mentioned mechanism, defects and epitaxial strain can be important factors [17-21]. Here, we report a controllable monolayer-critical magnetic effect, whereby a uniform ferromagnetic state appears in LMO at a critical thickness of 6 uc.

In our study, LMO (001) films with a thickness varying from 1 uc to 24 uc were grown using pulsed laser deposition monitored by reflection high energy electron diffraction (RHEED), on TiO$_2$-terminated (001)-oriented STO substrates, in a 10$^{-2}$ mbar oxygen pressure. LMO films are coherently strained up to 20 uc, as verified by X-ray diffraction and RHEED. All the films were found to be insulating by conductance measurement.

The distribution of local magnetic stray field emanating from the LMO films was imaged by scanning Superconducting QUantum Interference Device (SQUID) microscopy [22], called SSM, in zero applied field at 4.2 K. The essential part of the SSM is the square pickup loop with an inner size of ~ 3 x 5 μm (Fig. 1a). During the measurement, the pickup loop was scanned ~ 2 μm above the sample surface at a contact angle of approximately 10 degrees.

The SSM records the variation of magnetic flux threading the pickup loop and the flux detected by the pickup loop is converted to magnetic field by dividing by the effective pickup area of ~ 15 µm$^2$. The typical flux-sensitivity of the SSM is around 14 µ$\Phi_0$Hz$^{-1/2}$, where $\Phi_0$ = 2×10$^{-15}$ Tm$^2$ is the flux quantum and the bandwidth is 1000 Hz. Since our SSM sensor has a 10 degree inclination, the measured magnetic stray field component, denoted as $B_z$, is almost perpendicular the sample surface. The SSM images are x-y maps of the magnetic field-values (which are converted to a colour scale). The practical sensitivity during measurements was set by external noise sources, and is estimated to be about 30 nT.

Figure 1b shows a typical scan of a 200 x 200 µm area of a 6 uc LMO film grown on STO, with a pixel size of 1 x 1 µm. The cooling and measurement were both performed in zero applied field. An irregular pattern of regions with opposite magnetic field orientation was found, presenting a direct image of magnetic field emanating from ferromagnetic domains in the LMO. In Figs. 1c and 1d, the local magnitude variations along two orthogonal directions are presented. Bulk magnetization measurements of the 6 uc sample found a polarization of the film of 0.3 T, corresponding to a net moment of 1.6 µ$_B$ per Mn atom. Since the spin-only moment on Mn$^{3+}$ is 4 µ$_B$, the ferromagnetism appears to be noncollinear.

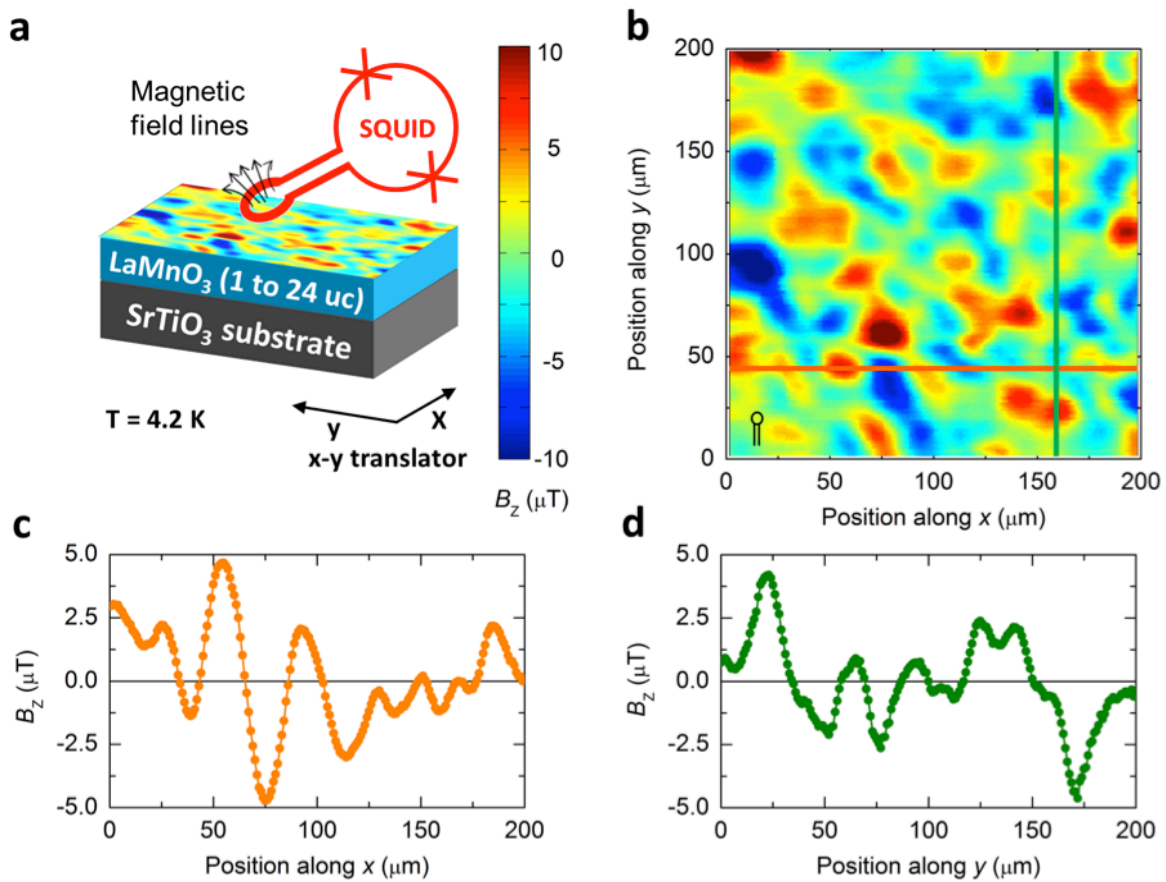

**Figure 1 | Scanning SQUID Microscopy (SSM) measurement on a 6 uc LMO film on a STO substrate. a**, Schematic drawing of the microscopy technique with sketch of pickup loop (red). **b**, Image of the inhomogeneous stray field distribution of a 6 uc LMO film at 4.2 K. The red-yellow peaks in the two-dimensional colour map indicates region where magnetic stray field exiting the sample, and the blue peak indicates region where magnetic stray field entering into the sample. The scan direction is always horizontally from left to right. The size of the pickup loop is indicated by a sketch (black) in the bottom-left corner of (b). **c,d**, *x*- (c) and *y*-direction (d) magnetic profiles for the corresponding lines in (b).

Figure 2 shows that there is a monolayer-sharp transition to ferromagnetism at a critical thickness of 6 uc. Figures 2 a-d indicate that the ferromagnetic domain size decreases with thickness above the critical value. The SSM signals for films with a thickness smaller than the critical thickness (Figs. 2e and 2f) are uniformly much weaker and cannot be resolved; they are two orders of magnitude smaller than the typical root-mean-square values of the thicker films. In the thinner samples, the isolated dipoles occasionally show up. In order to make a better visual contrast, we used the images with the dipoles. The critical thickness for ferromagnetism was confirmed by SSM measurement on another set of samples fabricated in a different growth chamber. Since uniform and controllable ferromagnetic state is necessary for device application, the observed critical thickness for ferromagnetism is of great significance. Huijben *et al.* [3] and Xia *et al.* [4] have reported the critical thickness for ferromagnetism in Sr doped LMO and $SrRuO_3$ thin films using bulk sensitive technique. However, due to the lack of domain structure information, characteristics and origin of the transition remain unclear. Using the SSM, Kalisky et al. [6] performed a comprehensive and detailed imaging study on magnetic structures in the interface between $LaAlO_3$ and STO below and above critical thickness for ferromagnetism, and they observed magnetic pitches above the critical thickness. However, there has been no report abrupt magnetic transition where the whole film switches to ferromagnetic phase yet.

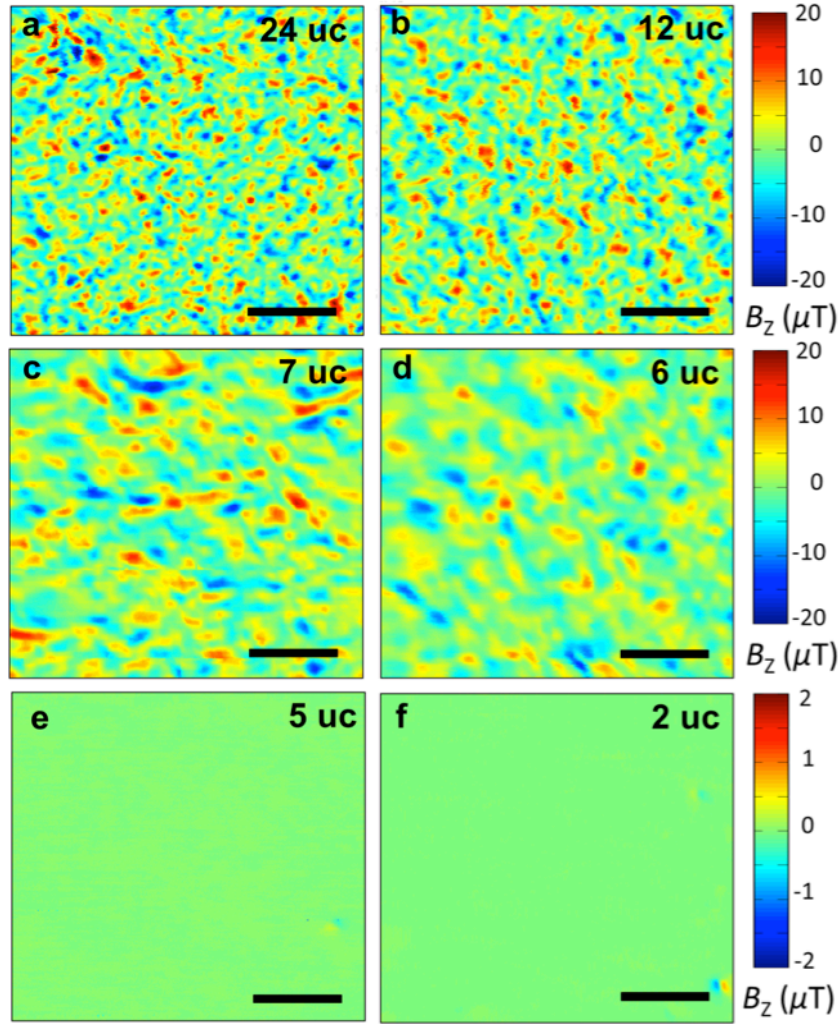

**Figure 2 | Critical thickness for ferromagnetism in insulating LMO (001) films grown on STO substrates. a-d**, Images of magnetic field emanating from LMO films with a thickness of (a) 24 uc, (b) 12 uc, (c) 7 uc and (d) 6 uc, respectively. **e,f**, Absence of magnetic field for 5 uc (e) and 2 uc (f) LMO. The scale bar corresponds to 100 µm. The scale of colour bars for 5 uc and 2 uc LMO are one order smaller than those of the other images.

To find out whether the ferromagnetism is indeed solely dependent on the LMO thickness, we fabricated and measured a sample that consisted partly of a 5 uc-thick and partly of a 7 uc-thick LMO film. The sample was prepared first by growing the 5 uc LMO film, then an *ex situ* a shadow mask was mounted to protect half of the sample, after which 2 extra uc of LMO were grown on the non-covered part. The sample was subsequently imaged by SSM following the usual measurement procedure. As seen from Fig. 3a, the 5 uc area does not display any signal within the noise level and the 7 uc area reveals an inhomogeneous magnetic field distribution comparable to the regular 7 uc LMO films (see Fig. 2c). Figure 3b demonstrates the line profile of the magnetic field-value difference between the 7 uc area and the 5 uc area. This experiment demonstrates the ability to modify the magnetic properties of this oxide compound by the controllable addition of less than 1 nm of the same material.

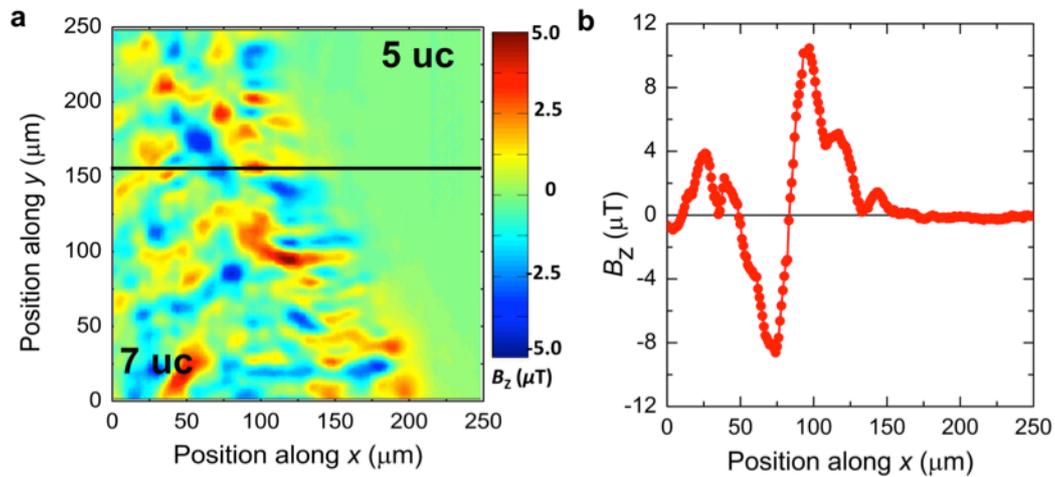

**Figure 3 | SSM images of a 5uc/7uc LMO film. a**, The upper-right area is the area where 5 uc LMO was grown and the bottom-left area is covered by 7 uc LMO. **b**, The *x*-direction magnetic profile for the black line in (a).

The magnetization orientation of the ferromagnetic LMO films was determined by the measuring magnetization of a 7 uc LMO film grown on STO along different orientations using vibrating sample magnetometer. During the measurement, the sample undergoes first a cooling process in 1 Tesla magnetic field, then the magnetic moments are measured during warm up in 0.1 Tesla. Figure 4a shows the temperature dependent magnetic moments of a 7 uc LMO film along two different orientations, revealing the in-plane nature of the magnetization. Therefore, the magnetic field pattern imaged by SSM is from the in-plane ferromagnetism. Magnetic moment as function of temperature was also measured for LMO films with different thicknesses using the same procedure. As shown in Fig. 4b, 4 uc and 5 uc LMO films do not show any clear net magnetic moment. Since the thinner films do not show ferromagnetic or Curie-law paramagnetic behaviour, the absence of magnetic field revealed by SSM in the films thinner than 6 uc is attributed to antiferromagnetic order. Furthermore, since there is no uncompensated magnetic flux observed in LMO films with odd number of layers, the spins are therefore perpendicular to the substrate.

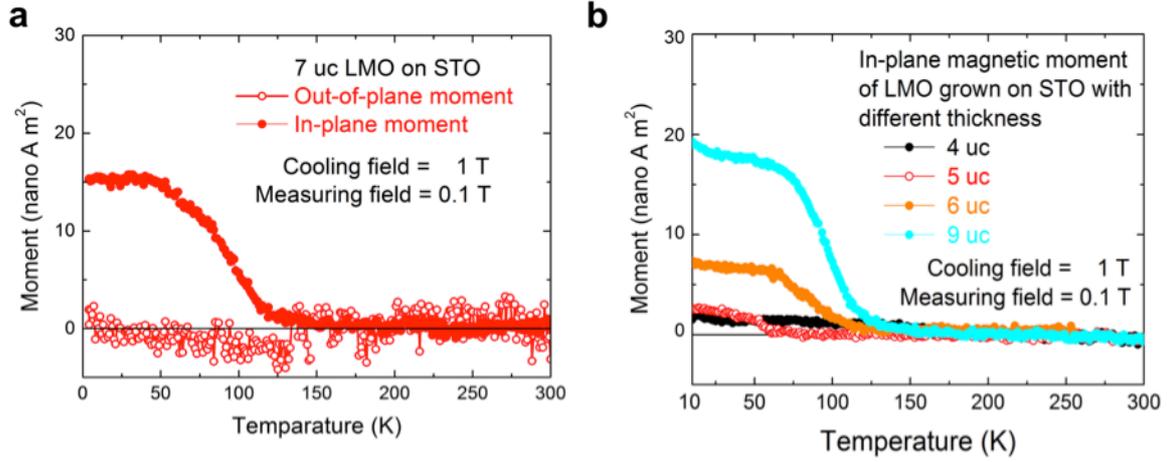

**Figure 4 | Bulk magnetic properties of LMO films grown on STO. a**, In-plane and out-of-plane magnetic moments of 7 uc LMO grown on STO. The magnetic moment of 7 uc of LMO is found to lie in-plane. **b**, Magnetic moment of 4 uc, 5 uc, 6 uc and 9 uc LMO films grown on STO as a function of temperature.

Because of the abruptness of the phase transition and the similarity in polar properties of LMO (001) and LaAlO$_3$ (001) films, it is relevant to consider possible electronic reconstruction, which has been proposed as the mechanism for the abrupt insulator-to-metal transition in the well-studied SrTiO$_3$-LaAlO$_3$ case [23]. LMO contains alternatively charged (LaO)$^{1+}$ and (MnO$_2$)$^{1-}$ layers resulting in an internal electric field $E_0$. A simple first order estimate for this field is $E_o = e/2A\varepsilon_o\varepsilon_r$; where $e$ is the electronic charge, $A$ is the unit cell area, $\varepsilon_o$ is the permittivity of vacuum and $\varepsilon_r$ is the dielectric constant of LMO. Taking $\varepsilon_r \sim 70$ at low temperature [24], the value of $E_0$ is 0.85 V/nm, which will shear both the valence and conduction bands, as shown in Fig. 5a. The band gap ($E_g$) in bulk LMO (about 1.3 eV [25]) is smaller than that in bulk STO (3.2 eV), and therefore charge transfer to eliminate the polar discontinuity occurs entirely within the polar LMO film. At a certain thickness $t_c = E_g/E_o \approx 4$ uc, the valence band maximum of the LMO reaches the conduction band minimum at the LMO/STO interface, initiating electron transfer from the top to the bottom of the LMO film. Such transfer then decreases the electric field in LMO. The transferred charge as a function of thickness is zero below $t_c$, and increases asymptotically with increasing thickness to $0.5e$ [26]. The electron transfer makes the interface region of LMO electron-doped and the top surface region of LMO hole-doped. This is analogous to the electronic phase separation known for bulk manganites [27], which in our case is stabilized by an intrinsic electric field of the polar LMO film. The doping of LMO tilts the exchange interaction between Mn ions from superexhange to double exchange and lead to ferromagnetism when the sufficient number of dopants is available. Our density-functional calculations predict that hole-doping favours the ferromagnetic ground state when the doping exceeds 0.08 e/Mn, in agreement with the experimentally reported bulk phase diagram of LMO [28,29]. Thus, it is expected that electronic reconstruction will favour ferromagnetism above a certain critical thickness of LMO. We note that, contrary to the LAO/STO system, where the interface becomes metallic above the critical thickness, our films remain insulating. This difference is due to the large band gap of LAO which leads to electron transfer to the conduction band of STO, whereas in the LMO/STO case the electronic reconstruction is revealed in the self-doping of LMO.

The exact thickness at which the ferromagnetism occurs depends on the spread of the charge. Figure 5b shows an estimated doping charge per Mn atom when all the charge is projected onto just one uc (black curve) and when the charge is spread over 2 uc (red curve). We see that in the latter case the bulk phase diagram predicts phase transition from the insulting antiferromagnetic phase to the insulating ferromagnetic phase when LMO thickness exceeds 6 uc, which is consistent with our experimental observations. The assumption of the charge spread over 2 uc agrees with results of first-principles calculations [30] that indicate screening of the interface charge in doped LMO.

In order to verify the feasibility of this electronic reconstruction mechanism, we compared the behaviour of 12 uc LMO and $CaMnO_3$ films, since the $Ca^{2+}Mn^{4+}O_3$ (001) film is non-polar. These films were grown on $TiO_2$-terminated conducting 0.1 wt% Nb-doped STO (Nb:STO) substrates and were covered by 2 uc $LaAlO_3$ under the same conditions. The 2 uc $LaAlO_3$ capping layer was used to reveal the effect of surface symmetry breaking, and the conducting Nb:STO substrate was used to investigate interface band-bending. By comparing SSM images between $LaAlO_3$ capped LMO grown on Nb:STO (Fig. 5b) and LMO grown on STO (Fig. 2c), we conclude that neither the surface symmetry breaking nor the band-bending effect contributes to the critical magnetic behaviour. Furthermore, as shown in Figs. 5c and 5d, only the LMO films show the magnetic field patterns, while the non-polar $CaMnO_3$ films do not reveal any signature of ferromagnetism. Thus, both our theoretical analysis and experimental data suggest that electronic reconstruction driven by the polar nature of LMO (001) films is the likely explanation for the abrupt transition to ferromagnetic order in LMO thin films.

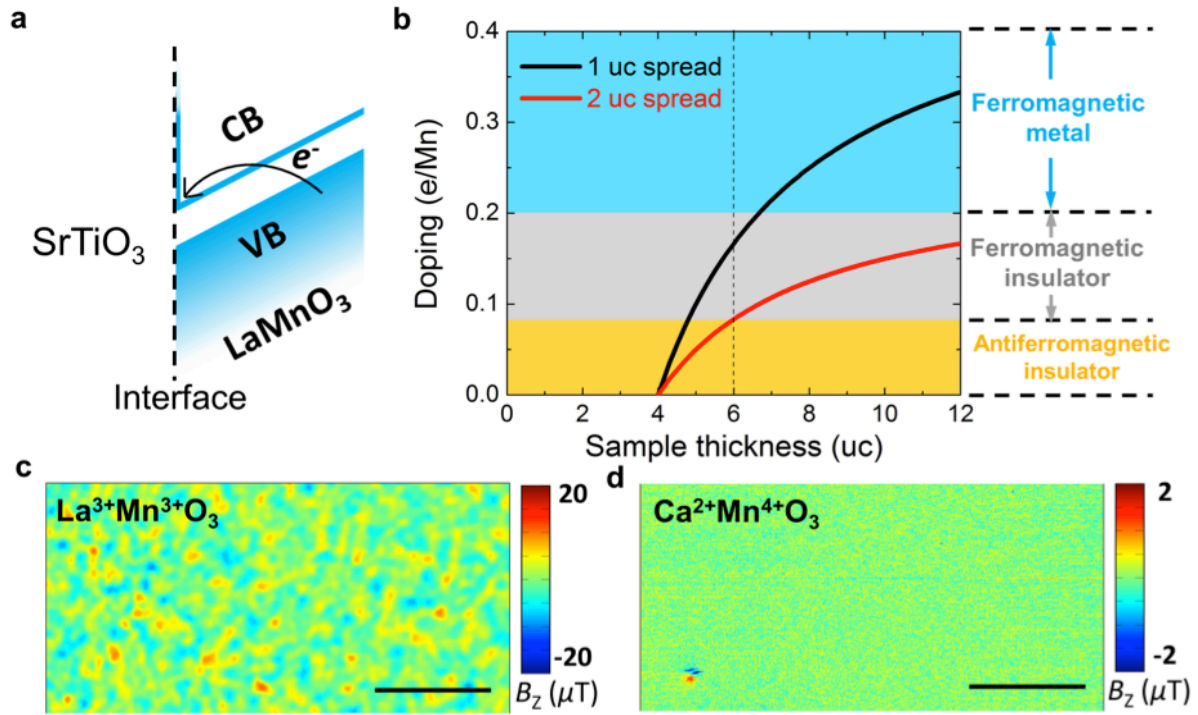

**Figure 5 | Analysis of the origin of ferromagnetism in LMO. a,** Band diagram of electronic reconstruction for LMO (001) film grown on an STO substrate. **b**, Amount of charge on the Mn site transferred from surface to interfacial layers as a function of LMO thickness. The black curve corresponds to the doping level where all the charge is projected onto just one uc and the red curve corresponds to the doping level when the charge is spread over 2 uc. The three doping regimes corresponding to ferromagnetic conducting state (blue area), ferromagnetic insulating state (gray area) and antiferromagnetic insulating state (orange area) are adapted from literature for doped bulk LMO. **c**, SSM image of 12 uc $La^{3+}Mn^{3+}O_3$ showing ferromagnetic behaviour. **d**, SSM image of 12 uc $Ca^{2+}Mn^{4+}O_3$ shows no sign of ferromagnetic behaviour. The scale bar corresponds to 100 μm.

Summarizing, we have imaged the distribution of the magnetic stray field emanating from insulating LMO films grown on STO substrates. An atomically sharp antiferromagnetic-to-ferromagnetic transition was revealed when the LMO film thickness is increased to 6 uc. This critical thickness for ferromagnetism is explained in terms of electronic reconstruction, involving electron transfer from the top of the LMO film to the bottom. Our results demonstrate how emergent functionality, in this case ferromagnetism, can be visualized and engineered at the atomic level.

**Acknowledgements**
We thank Z. L. Liao, M. S. Golden, A. Brinkman, D. Schlom, M. Huijben, Z. L. Liao, G. Campi and H. M. Christen for valuable discussions and D. Veldhuis for assistance in SQUID design and fabrication, J. R. Kirtley for providing the Scanning SQUID and Microscopy setup and the IBM T. J. Watson Research Centre. This research was financially supported by the Dutch FOM and NWO foundations through a VICI grant and a Rubicon grant (2011, 680-50-1114). The work at National University of Singapore is supported by the Singapore National

Research Foundation (NRF) under the Competitive Research Programs (CRP) "Tailoring Oxide Electronics by Atomic Control" (CRP Award No. NRF2008NRF-CRP002-024) and "New Approach To Low Power Information Storage: Electric-Field Controlled Magnetic Memories" (CRP Award No. NRF-CRP10-2012-02). Parts of this research were carried out at the Stanford Synchrotron Radiation Lightsource, a Directorate of SLAC National Accelerator Laboratory and an Office of Science User Facility operated for the U.S. Department of Energy Office of Science by Stanford University. The work at University of Nebraska was supported by the National Science Foundation through Materials Research Science and Engineering Center (MRSEC, Grant No. DMR-0820521). J.M.D.C. acknowledges support from Science Foundation Ireland Grant 10/IN1.13006 and from the EU FP7 IFOX Project.

**Author contributions**
X.R.W. performed the SQUID measurements. X.R.W., M. H. and H.H. analysed the results. C.J.L., W.M.L., T.V. and A. fabricated the samples. T.R.P. and E.Y.T. performed the first-principles calculations. A.V., D.P.L., M.H. and N.P. performed X-ray diffraction measurements. X.R.W, C.J.L J.M.D.C and W.M.L performed bulk magnetization measurement and analysis. X.R.W. W.M.L. and H.H. guided the work. X.R.W. and H.H. wrote the manuscript with input from J.M.D.C and the other co-authors.